\documentclass[twocolumn,superscriptaddress,amsfont,amssymb,amsmath,showpacs,balancelastpage,nofootinbib,aps,pra,10pt]{revtex4-2}
\usepackage{graphicx,longtable,mathrsfs,color,array}
\usepackage[unicode=true,pdfusetitle,
bookmarks=true,bookmarksnumbered=true,bookmarksopen=true,bookmarksopenlevel=1,
breaklinks=false,pdfborder={0 0 0},backref=false,colorlinks=true]{hyperref}
\hypersetup{citecolor=blue,filecolor=blue,linkcolor=blue,urlcolor=blue,pdfauthor={Name}
}
\usepackage[dvipsnames,table]{xcolor}
\usepackage{amssymb,amsmath,mathtools,mathrsfs,enumitem}
\usepackage{epsfig,subfigure,placeins,float}
\usepackage{booktabs,multirow}
\usepackage{exscale,relsize}
\usepackage[normalem]{ulem}
\usepackage[T1]{fontenc}
\usepackage[utf8]{inputenc}
\usepackage{enumerate}
\usepackage{times, mathptmx}
\usepackage{tikz}
\usepackage{aas_macros}
\usepackage{tabularx}
\usepackage{tabularray}
\usepackage{diagbox}
\usepackage{multirow}
\usepackage{pifont}
    \newcommand{\cmark}{\ding{51}}
    \newcommand{\xmark}{\ding{55}}
\newcolumntype{M}[1]{>{\centering\arraybackslash}m{#1}}
\usepackage{hhline}
\usepackage{makecell}
\usepackage{cleveref}
\usepackage{siunitx}
\usepackage{multirow}
\usepackage{booktabs}

\newcolumntype{Y}{>{\centering\arraybackslash}X}

\usetikzlibrary{arrows,positioning,decorations.markings,decorations.pathmorphing,calc}

\newcommand{\be}{\begin{equation}}
\newcommand{\ee}{\end{equation}}
\newcommand{\ba}{\begin{align}}
\newcommand{\ea}{\end{align}}

\newcommand{\comment}[1]{}

\newcolumntype{C}[1]{>{\centering\let\newline\\\arraybackslash\hspace{0pt}}m{#1}}

\newcommand{\contwo}{c_2}
\newcommand{\conthree}{c_3}
\newcommand{\approxcon}{\beta}

\newcommand{\nocontentsline}[3]{}
\newcommand{\tocless}[2]{\bgroup\let\addcontentsline=\nocontentsline#1{#2}\egroup}

\def\mpl{M_{\rm Pl}}
\def\Mpl{M_{\rm Pl}}

\newcommand{\jnh}[1]{}

\definecolor{hyperref}{RGB}{026,028,087}

\def\gsim{ \lower .75ex \hbox{$\sim$} \llap{\raise .27ex \hbox{$>$}} }
\def\lsim{ \lower .75ex \hbox{$\sim$} \llap{\raise .27ex \hbox{$<$}} }

\newlength{\stheight}
\newcommand\textst[1][fu-grey]{
\ifmmode\setlength{\stheight}{+1.0ex}
\else\setlength{\stheight}{+0.5ex}
\fi
\bgroup\markoverwith{\textcolor{#1}{\rule[\the\stheight]{2pt}{1.0pt}}}\ULon
} 
 
\newcommand{\textins}[2][fu-grey]{
\ifmmode\mathcolor{#1}{#2}
\else\textcolor{#1}{#2}\@\,
\fi
}
\graphicspath{{./Plots/}}
\allowdisplaybreaks

\tikzstyle{vecArrow} = [thick, decoration={markings,mark=at position
1 with {\arrow[semithick]{open triangle 60}}},
double distance=1.4pt, shorten >= 5.5pt,
preaction = {decorate},
postaction = {draw,line width=1.4pt, white,shorten >= 4.5pt}]

\newcommand{\order}[1]{{\cal O}\left(#1\right)}

\begin{document}

\setcounter{tocdepth}{5}
\title{Testing Dark Energy with Black Hole Ringdown}

\author{Laurens Smulders}
\affiliation{Department of Physics \& Astronomy, University College London, London, WC1E 6BT, U.K}

\author{Johannes Noller}
\affiliation{Department of Physics \& Astronomy, University College London, London, WC1E 6BT, U.K}
\affiliation{Institute of Cosmology \& Gravitation, University of Portsmouth, Portsmouth, PO1 3FX, U.K.}

\author{Sergi Sirera}
\affiliation{Institute of Cosmology \& Gravitation, University of Portsmouth, Portsmouth, PO1 3FX, U.K.}
\affiliation{Department of Physics \& Astronomy, University College London, London, WC1E 6BT, U.K}

\begin{abstract}
We show that dynamical dark energy theories can imprint ${\cal O}(1)$ modifications on the quasi-normal mode (QNM) spectrum characterising black hole ringdown. The time dependence of dynamical dark energy naturally gives rise to cosmological `hair' around a black hole. Taking the cubic Galileon as a concrete example -- which admits the only known stable solution of this kind -- we parametrically connect the cosmological and black hole regimes, derive the induced QNM shifts and forecast the resulting dark energy constraints. We find that the dark energy field profile can be constrained with an accuracy of up to $10^{-2}$ for LVK and $10^{-4}$ for LISA.
\end{abstract}

\date{\today}
\maketitle

\section{Introduction} \label{sec:Intro}

The remarkable simplicity of black holes in General Relativity (GR) \cite{Israel:1967wq,Carter:1971zc,Bekenstein:1971hc, Ruffini:1971bza, Bekenstein:1972ky,Robinson:1975bv} makes them a powerful laboratory for detecting new gravitational physics. Cosmology provides some of the most compelling theoretical and observational clues suggesting that physics beyond GR -- or, in a cosmological context, beyond $\Lambda{}$CDM -- may be present, see e.g. the recent \cite{DESI:2024mwx,DES:2025sig}. In particular, the presence of dynamical dark energy, i.e. dark energy different from a pure cosmological constant $\Lambda$, would generically introduce new degrees of freedom with non-trivial time dependence -- see \cite{Copeland:2006wr,Clifton:2011jh,Babichev:2013vji,Joyce:2014kja,Koyama:2015vza,Kobayashi:2019hrl} for reviews. This time dependence naturally gives rise to cosmological hair in the context of black hole solutions, i.e. the local black hole physics is no longer controlled by just the mass, spin and charge of the black hole (as in GR/$\Lambda{}$CDM), but also impacted by the dynamical dark energy field, see e.g. \cite{Jacobson:1999vr,Babichev:2013cya}. Yet, until recently, when their stability was fully investigated, all such known solutions were found to be unstable \cite{Babichev:2013cya,Kobayashi:2014eva,Babichev:2016kdt,BenAchour:2018dap,Motohashi:2019sen,Charmousis:2019vnf,deRham:2019gha,Takahashi:2020hso,Khoury:2020aya,Takahashi:2021bml}. This changed with the recent discovery of a stable solution in \cite{Smulders:2026qwc}. Derived in the context of the cubic Galileon \cite{Nicolis:2008in} -- an illustrative theory sharing key characteristics of much wider classes of scalar-tensor theories -- this solution recovers the desired cosmological long-range behaviour and gives rise to well-behaved short-range dynamics around black holes. In this paper we show how the cosmological and black hole regimes are parametrically linked for this solution, opening the door for a {\it direct} test of cosmological physics with the local emission processes probed by black hole ringdown. We compute the quasi-normal mode (QNM) spectrum, providing a {\it smoking gun signature} for dynamical dark energy in black hole ringdown, and {\it forecast} the precision with which these observables will be testable by current and future experiments. Throughout we work in natural units, where $c=1 =\hbar$.

\section{Stable Black Holes with cosmological hair} \label{sec:stableBH}

Black hole solutions in scalar-tensor theories with a time-dependent scalar have primarily been investigated in the context of stealth metrics (i.e. solutions identical to those in GR, e.g. Schwarzschild) and linearly time-dependent scalar field solutions (a choice motivated by simplicity and the fact that, in shift-symmetric theories, it helps derive static solutions). In Table \ref{tab:lit-rev} we summarise such black hole solutions alongside other known examples going beyond exact stealth solutions. The non-static cubic Galileon solution stands out as the only known stable solution, tantalisingly suggesting that either going beyond staticity and/or beyond exact stealth solutions may be required to obtain stable solutions in typical scalar-tensor theories.\footnote{
Closely related to this, \cite{DeFelice:2022xvq} shows that strong coupling problems in DHOST theories can be avoided by including a \textit{scordatura} term \cite{Motohashi:2019ymr}. In \cite{DeFelice:2022qaz} it is then shown that exact stealth solutions for such a theory do not exist, but approximate ones with a small time dependence do. Given known instability results \cite{Babichev:2013cya,Kobayashi:2014eva,Babichev:2016kdt,BenAchour:2018dap,Motohashi:2019sen,Charmousis:2019vnf,deRham:2019gha,Takahashi:2020hso,Khoury:2020aya,Takahashi:2021bml} typically rely on assuming solutions with $X=\text{const}$, a non-constant $X$ might also be required for stability. Note that we here focus on solutions in specific covariant theories, but refer to \cite{Franciolini:2018uyq,Noller:2019chl,Hui:2021cpm,Mukohyama:2022enj,Khoury:2022zor,Mukohyama:2022skk,Mukohyama:2023xyf,Mukohyama:2024pqe,Barura:2024uog,Mukohyama:2025jzk,Mukohyama:2025owu} for recent work studying black-hole perturbations using effective field theory techniques (detached from the question of whether a given perturbative solution can
be embedded in a full covariant theory). 
}
\\

\begin{table*}[t]
\centering
\begin{tabular*}{0.8\textwidth}{@{\extracolsep{\fill}} l l l @{}}
    \toprule
    \textbf{Theory} & \textbf{Background solution} $\{g_{\mu\nu}, X \equiv -\tfrac{1}{2}\partial_\mu\phi\partial^\mu\phi \}$ & \textbf{Stability} \\
    \midrule

    (Shift + refl)-sym Horndeski  & S(dS) metric, $X = q^2/2 = \text{const}$ \cite{Mukohyama:2005rw,Babichev:2013cya,Kobayashi:2014eva} & \xmark \cite{Takahashi:2016dnv, Takahashi:2021bml} \\ \arrayrulecolor{gray}\midrule\arrayrulecolor{black}

    \multirow{2}{*}{Shift-sym quadratic DHOST} & $\text{S(dS)}+\text{K}$ metric, $X = \text{const}$ \cite{Motohashi:2019sen,Charmousis:2019vnf} & \xmark \cite{Takahashi:2019oxz, deRham:2019gha,Takahashi:2021bml} \\ 
    & $\text{S(dS)}+\text{(K)RN(dS)}$ metric, $X = \text{const}$ \cite{Takahashi:2020hso} & \xmark \cite{Takahashi:2019oxz, deRham:2019gha,Takahashi:2021bml} \\ \arrayrulecolor{gray}\midrule\arrayrulecolor{black}
    
    \multirow{2}{*}{Cubic Galileon} & Hairy static metric, non-const $X$ \cite{Babichev:2012re,Babichev:2016fbg} & \xmark \cite{Smulders:2026qwc} \\
     & Hairy non-static metric, non-const $X$ \cite{Babichev:2025ric,Smulders:2026qwc} & \cmark \cite{Smulders:2026qwc} \\
    
    \bottomrule
\end{tabular*}
\caption{
Known black hole solutions in scalar-tensor theories with a linearly time-dependent scalar, for which the stability of perturbations has been investigated comprehensively (i.e. for both odd and even parity modes). The hairy non-static solution for the Cubic Galileon stands out as the only known fully stable solution, while the other listed solutions suffer from instabilities in the even sector.
Other known hairy solutions, for which stability has yet to be fully investigated, include those proposed in \cite{Babichev:2016kdt,BenAchour:2018dap,Bakopoulos:2023fmv}, for which odd parity stability has been shown in \cite{Sirera:2024ghv,Kobayashi:2025evr,Charmousis:2025xug}, but even-parity stability remains to be assessed. Additionally, partial stability -- so-called `proto-stability' -- has been demonstrated (including even modes) for the hairy solutions found in \cite{Lara:2025hqh}.
This table builds on those of \cite{Motohashi:2019sen,Sirera:2024ghv}.
}
\label{tab:lit-rev}
\end{table*}

{\it Cubic Galileon}: Galileon scalar-tensor theories \cite{Nicolis:2008in} have been a key `lamppost' theory to illustrate the behaviour of large classes of scalar-tensor theories, both in cosmological and strong gravity regimes. Crucial features exemplified by Galileons in particular are: 1) their non-linear interactions giving rise to self-accelerating solutions on larger scales \cite{Silva:2009km,DeFelice:2010pv} 2) a screening mechanism on smaller scales \cite{Nicolis:2008in,Burrage:2010rs,Babichev:2013usa}, 3) their shift symmetry resulting in analytically tractable cosmological attractor solutions \cite{DeFelice:2010pv}. The action for the cubic Galileon, the simplest such example, is given by
\begin{align}
    \label{eq:CubicGalileonAction}
    S=\int d^4x\sqrt{-g}\left[
        \frac{\mpl^2}{2}R
        +\frac{1}{2} (\partial\varphi)^2
        -\frac{\conthree}{\Lambda_3^3}\Box\varphi(\partial\varphi)^2
    \right],
\end{align}
where in addition to the scalar $\varphi$, we have included the standard kinetic term for the metric $g_{\mu\nu}$, the Ricci scalar $R$. $g$ is the determinant of the metric, $\mpl$ is the reduced Planck mass, and $\Box \equiv \nabla_\mu\nabla^\mu$ is the d'Alembertian operator. $\conthree$ is a dimensionless parameter and we impose $\Lambda_3^3 = \mpl H_0^2$, choosing a dark energy motivated mass scale for $\Lambda_3$ that ensures an ${\cal O}(1)$ contribution to the Friedmann equations, where $H_0$ is the value taken by the Hubble constant today. Note that we have fixed the normalisation of the Galileon field $\varphi$ to give the $+\frac{1}{2}$ coefficient in front of the kinetic term, where the sign is mandated by requiring the existence of viable cosmological self-accelerating solutions \cite{DeFelice:2011bh,Barreira:2013jma,Melville:2022ykg}, i.e. when treating this model as dynamical dark energy. 
\\

{\it Stable black hole solutions}: As shown in Table~\ref{tab:lit-rev}, most known black hole solutions in candidate (scalar-tensor) dynamical dark energy theories -- where the dark energy field is therefore naturally time-dependent -- are unstable. In the absence of identifying a stabilising mechanism, no physical ringdown predictions can therefore be extracted for such models. However, recently, stable black hole solutions with cosmological (dark energy) hair were found for the cubic Galileon in \cite{Smulders:2026qwc}.\footnote{
Specifically, the ghost and gradient stability of these solutions was fully investigated in the vicinity of the black hole while the stability on the cosmological asymptotes was previously shown in \cite{Kobayashi:2009wr,DeFelice:2011bh}.}
This opens the door for computing physical ringdown predictions in the presence of dynamical dark energy, so we here summarise the key aspects of this stable background solution relevant for this computation. The metric and scalar take the general spherically symmetric form
\begin{align}
    \label{eq:Ansatz}
    ds^2&=-h(t,r)dt^2+\frac{dr^2}{f(t,r)}+r^2d\Omega^2,
    \nonumber\\
    \varphi&=qt+q\Psi(t,r),
\end{align}
where $h$, $f$ and $\Psi$ are functions of time $t$ and radius $r$, $d\Omega^2$ is the angular line element and $q$ is a constant related to the accelerated expansion on large scales.

No exact black hole solutions are known for the ansatz \eqref{eq:Ansatz}, but a cosmological (de Sitter-like) long range limit with Hubble parameter $H^4= \Lambda_3^6/(216c_3^2\mpl^2)$ is readily obtained. In order for the cosmological limit to be homogeneous, i.e. for the inhomogeneities due to the presence of the black hole to decay on length scales $\ll H^{-1}$, we require
\begin{align}
    \label{eq:qBound}
    \frac{q^2-q_0^2}{q_0^2}\ll 1,
    \quad \text{where} \quad
    q_0=-\text{sign}(\conthree)\sqrt{6}\mpl H
\end{align}
Finally, a highly accurate\footnote{
As shown in \cite{Smulders:2026qwc}, the solutions \eqref{eq:ApproximateSolutions} are accurate as long as $r\ll H^{-2/3}r_s^{1/3}$.
}
, stable short-range solution  near the black hole (using $H r_s \ll 1$) can be obtained
\begin{align}
    \label{eq:ApproximateSolutions}
    h(t,r)&=\approxcon f(t,r)=\approxcon \left(1-\frac{r_s(t)}{r}\right),
    \nonumber\\
    \Psi(t,r)&=\frac{r_s(t)}{\sqrt{\approxcon}}\ln\left|\frac{\sqrt{u}-1}{\sqrt{u}+1}\left(\frac{\sqrt{u}+2}{\sqrt{u}-2}\right)^{\frac{1}{2}}\right|,
    \quad
    u \equiv 4 - \frac{3 r_s(t)}{r},
    \nonumber\\
    r_s(t)&=r_s(0)\exp\left[\left(\frac{q}{q_0}\right)^3 \frac{H t}{\approxcon}\right],
\end{align}
where $\approxcon$ is an $\order{1}$ integration constant and together with $q$ encodes the black hole `hair'. This is the background relevant for the emission of gravitational waves and so will be crucial to compute quasi-normal mode spectra later. Note that $\approxcon$ can be removed {\it locally} by rescaling $t\rightarrow t/\sqrt{\approxcon}, q \rightarrow q\sqrt{\beta}$, which means that $\approxcon$ cannot be determined using local observables only.\footnote{We thank Shinji Mukohyama for discussions on this point.} However, the expansion rate of the universe on large scales defines a reference time-normalisation, making $\approxcon$ an observable parameter when considering the embedding of the black hole in an expanding universe. Lastly, equation \eqref{eq:ApproximateSolutions} shows that the black hole evolves on timescales $H^{-1}$, much longer than other relevant black hole evolution timescales, so these solutions are \textit{quasi-stationary}. Likewise, $\Psi$ in \eqref{eq:Ansatz} also evolves on timescales $\sim H^{-1}$, so $\varphi$ approximately retains linear time dependence, like other solutions shown in table \ref{tab:lit-rev}.
 
\section{Connecting cosmological and black hole solutions} \label{sec:FittingFormula}

While $\approxcon$ is just an integration constant in the context of the short-scale solution \eqref{eq:ApproximateSolutions}, matching this to the desired cosmological large-scale behaviour using the full solution holds the key to parametrically understanding the black hole hair in terms of dark energy dynamics. In the absence of a known analytical solution for all scales, numerical integration of the full equations of motion -- see Appendix~\ref{sec:AppendixEOMs} -- can be used to link these scales -- see \cite{Smulders:2026qwc} for details on the numerical procedure. 

\begin{figure}[t]
    \centering
    \includegraphics[width=\linewidth]{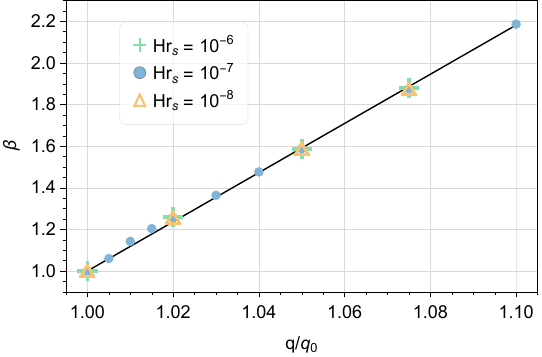}
    \caption{
    Values for the hair parameter $\approxcon$ obtained using the \textit{shooting method} \cite{Babichev:2016fbg,Emond:2019myx,Smulders:2026qwc} for a range of scalar velocities $q/q_0$ and example $Hr_s$ values. We see an excellent fit with a linear dependence on $q/q_0$ given by equation \eqref{eq:BetaFittingFormula}, and no leading order dependence on $Hr_s$.
    }
    \label{fig:BetaFittingFormula}
\end{figure}

After substituting in the ansatz \eqref{eq:Ansatz}, the equations of motion are controlled by the parameters $H$, $q/q_0$ and $r_s$ (which enters via defining boundary conditions). These three parameters therefore fully describe the solutions and hence also uniquely determine $\approxcon$. Since $\approxcon$ is dimensionless we must have $\approxcon=\approxcon(q/q_0,Hr_s)$. Given $Hr_s \ll1$ and $\approxcon\sim\order{1}$, it is natural to expect that the dependence on $H r_s$ is subleading\footnote{
This is the case if the functional dependence is polynomial, since there are no other parameters of the same order of magnitude in the problem. However, for a more complicated dependence (e.g. logarithmic) $Hr_s$ could in principle contribute to $\approxcon$ at the leading order.
}
and hence that $\approxcon=\approxcon(q/q_0)$. To verify this and find the full solution, we numerically integrate, applying a \textit{shooting} method \cite{Babichev:2016fbg,Emond:2019myx,Smulders:2026qwc} to determine the value of $\approxcon$ for a range of $Hr_s$ and $q/q_0$ values satisfying \eqref{eq:qBound}.\footnote{
Empirically, we could not find solutions approaching the cosmological asymptotes at large $r$ for $q<q_0$. More specifically, for every $q/q_0$, below a minimum value of $\beta$ no solutions seem to be obtainable due to a singularity appearing at an intermediate $r$. For all cases explored with $q<q_0$, we find that the value for $\beta$ required to approach the cosmological limit at large radii seems to lie below this minimum, and is therefore not reachable.
}
The results are plotted in Figure~\ref{fig:BetaFittingFormula}, confirming no leading order dependence on $Hr_s$. Applying a least-squares fit on the data points, we find excellent agreement with the following linear dependence on $q/q_0$:
\begin{align}
    \label{eq:BetaFittingFormula}
    \approxcon=11.82 \left(\frac{q}{q_0}-1\right)+1.
\end{align}
Note that, following \cite{Babichev:2016fbg,Emond:2019myx,Smulders:2026qwc}, we choose illustrative example values for $H r_s$ that are less extreme than the physical hierarchies (e.g. $H r_s \sim 10^{-17}$ for ${\cal O}(10^6) M_\odot$ black holes observable by LISA), in order to enable the computation of numerical solutions. The absence of a $H r_s$-dependence in \eqref{eq:BetaFittingFormula} then suggests that this result can be safely extrapolated to such more extreme hierarchies.

\section{\label{sec:QNMs}Quasi-normal mode spectra}

{\it Black hole perturbation theory}: To compute the quasi-normal mode (QNM) spectrum, we perturb the metric and scalar field as
\begin{align}
    g_{\mu\nu}=\bar{g}_{\mu\nu}+\frac{r_s}{\mpl}\delta g_{\mu\nu},
    \quad
    \varphi=\bar{\varphi}+\sqrt{\frac{\Lambda_3^3}{|c_3 q|r_s}}\delta\varphi,
\end{align}
where $\bar{g}_{\mu\nu}$ and $\bar{\varphi}$ are the background metric and scalar fields, while $\delta g_{\mu\nu}$ and $\delta \varphi$ are their perturbations, respectively.\footnote{
The normalisation is chosen such that the kinetic terms in the quadratic action found using these perturbations are canonically normalised with coefficients of $\order{1}$. Note that the factor of $r_s$ in front of the metric perturbation is a result of considering the normalisation in dimensionless coordinates $y=t/r_s$ and $x=r/r_s$ -- see \cite{Smulders:2026qwc} for further details.
}
We substitute these into the cubic Galileon action \eqref{eq:CubicGalileonAction} and expand to second order in perturbations to find the \textit{quadratic action}, working in \textit{Regge-Wheeler gauge} \cite{ReggeWheeler,Maggiore:2018sht} -- see Appendix~\ref{sec:AppendixRWGauge}.

The QNM spectrum is determined by the {\it local} dynamics of the black hole and its perturbations, so we work with the short-range solutions \eqref{eq:ApproximateSolutions} that are highly accurate in the vicinity of the black hole, substituting them into the quadratic action and working to leading order in $H r_s$, as before.\footnote{
Note that the WKB method for computing QNM frequencies \cite{Schutz:1985km,Konoplya:2019hlu} nicely illustrates the local `generation' of these modes, with all quantities in the calculation (the potential and its derivatives) evaluated at the maximum in the Regge-Wheeler/Zerilli potential, which for GR is at $\frac{3}{2} r_s$.
}
At leading order in $H r_s$, we can ignore the suppressed time dependence of the background solutions. Additionally, scalar and metric perturbations decouple and we therefore obtain a schematic decoupled quadratic action
\begin{align}
    \label{eq:QuadraticActionSchematic}
    \mathcal{L}_\text{leading order}^{(2)}
    =\mathcal{L}_\text{odd}^{(2)}
    +\mathcal{L}_\text{even}^{(2)}
    +\mathcal{L}_\text{scalar}^{(2)}.
\end{align}
To describe the ringdown signal in a gravitational wave detector, we are only interested in the metric parts. These are conveniently expressed in terms of the summary variables $w_{\ell m}$ and $\psi_{\ell m}$ as -- see Appendix~\ref{sec:AppendixRWGauge} for details --
\begin{widetext}
\begin{align}
    \label{eq:QuadraticActionsFull}
    \frac{4\sqrt{\approxcon}(\ell+2)(\ell-1)r_s^2}{\ell(\ell+1)}\mathcal L_{\text{odd}}^{(2)}
    &=\frac{r^3}{\approxcon(r-r_s)}\dot{w}^2
    -r(r-r_s)w'^2
    -\frac{\ell(\ell+1)r-4r_s}{r}w^2,
    \nonumber\\\nonumber\\
    \frac{\sqrt{\approxcon}\ell(\ell+1)r_s^2}{(\ell+2)(\ell-1)}\mathcal L_{\text{even}}^{(2)}
    &=\frac{r^3(r-r_s)}{\left((\ell+2)(\ell-1)r+3r_s\right)^2}\dot{\psi}^2
    -\approxcon\frac{r(r-r_s)^3}{\left((\ell+2)(\ell-1)r+3r_s\right)^2}\psi'^2
    \nonumber\\\nonumber\\
    &-\approxcon\frac{(r-r_s)\left(3r_s^2-\left(\ell(\ell+1)+4\right)r_s r+\ell(\ell+1)(\ell+2)(\ell-1)r(r-r_s)\right)}{\left((\ell+2)(\ell-1)r+3r_s\right)^3}\psi^2,
\end{align}
\end{widetext}
where we have omitted the summation over the spherical harmonic indices $(\ell, m)$. These expression are valid for $\ell\geq2$. As the metric does not possess monopole and dipole degrees of freedom, the $\ell=0$ and $\ell=1$ sectors are of no interest to us here.
\\

\begin{figure}[t!]
    \centering
    \includegraphics[width=\linewidth]{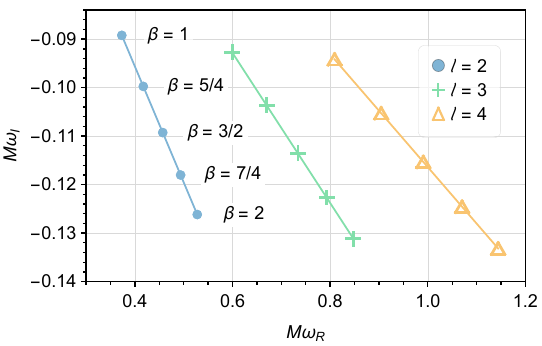}
    \caption{
    The quasi-normal mode (QNM) frequencies for different values of $\approxcon$, shown for $\ell = 2,3,4$. M is the (remnant) black hole mass and $\omega_I$ and $\omega_R$ are the imaginary and real parts of the complex QNM frequency $\omega$. Cosmological hair induced deviations from the GR limit $\approxcon = 1$ leads to QNM frequency modifications of up to $\sim 40 \; \%$. Here, using \eqref{eq:BetaFittingFormula}, $\approxcon\sim2$ is the maximum $\approxcon$ for which $q$ satisfies the bound formulated in \eqref{eq:qBound}, ensuring homogeneity on large scales. 
    }
    \label{fig:QNMShifts}
\end{figure}

{\it Modified Regge-Wheeler and Zerilli equations}:
We derive the equations of motion from the quadratic actions in Eq.~\eqref{eq:QuadraticActionsFull} and cast them into a Schr\"odinger-like form. We define the \textit{Regge-Wheeler} and \textit{Zerilli} functions respectively as:
\begin{align}
    Q(t,r)&=rw(t,r), 
    &Z(t,r)&=\frac{r(r-r_s)}{2\lambda r+3r_s}\zeta(t,r).
\end{align}
Introducing the \textit{tortoise coordinate} $r_*$ with $dr_*=dr/A(r)$, and $A(r)=\left(1-r_s/r\right)$, the equations of motion in Fourier space are
\begin{align}
    \label{eq:ModifiedRWZ}
    \left(\frac{\omega^2}{\approxcon}+\partial_{*}^2-V_\ell^{\rm RW,GR}(r)\right)Q(t,r)&=0, \nonumber \\
    \left(\frac{\omega^2}{\approxcon}+\partial_{*}^2-V_\ell^{\rm Z, GR}(r)\right)Z(t,r)&=0,
\end{align}
where $\partial_*\equiv \partial/\partial r_*$. $V_\ell^{\rm RW, GR}$ and $V_\ell^{\rm Z, GR}$ are the same Regge-Wheeler and Zerilli potentials as in GR, see Appendix~\ref{app:RWZ} for details. From this it follows that 
\begin{align}
    \label{eq:FrequencyShift}
    \omega=\sqrt{\approxcon}\omega_{GR},
\end{align}
where we note that the isospectrality of even and odd spectra is inherited from GR. The {\it smoking gun effect} of a cosmological cubic Galileon on the QNM spectrum is therefore the rescaling of all frequencies and decay times by $\sqrt{\approxcon}$. This is illustrated in Figure~\ref{fig:QNMShifts}. QNM measurements can therefore directly constrain the black hole hair (parametrised by $\beta$) in such setups, and through \eqref{eq:BetaFittingFormula} constrain $q/q_0$, thus enabling the reconstruction of the full scalar profile $\varphi$.

It is important to note that the frequency-shift induced by $\beta$ is entirely {\it degenerate} with a change in the black hole mass $M$ -- a consequence of the fact that $\beta$ can be removed locally by a time rescaling, as discussed in section \ref{sec:stableBH}. Taking the quadrupolar $\ell=2, n=0$ mode\footnote{
Note that, to linear order, for the background considered here, different modes do not mix. Moreover, non-rotating backgrounds result in $m$-independent equations of motion~\cite{Tattersall:2017erk}. 
} 
as an example, the numerical GR solution \cite{Chandrasekhar:1975zza} modified by the the $\beta$ correction yields
\begin{equation}
      \frac{M}{\sqrt{\beta}}\omega_{20} \approx 0.37367 - 0.08896i.
      \label{eq:qnm}
\end{equation}
This demonstrates that the effect of $\beta$ can be mimicked by scaling the mass $M \to M/\sqrt{\beta}$. While measuring multiple QNMs cannot break this degeneracy, additional information on the mass $M$ (e.g. derived from the inspiral or merger phase) can -- this could e.g. be obtained by adapting the closely related inspiral and merger calculations/simulations in \cite{deRham:2012fw,Dar:2018dra,Brax:2020ujo,Figueras:2021abd} to the solutions considered here.

\section{Forecasted Constraints and testing dark energy} \label{sec:Forecasts}

To estimate how precisely current and upcoming GW observations can constrain $\approxcon$, we perform a Fisher information analysis, as similarly done in~\cite{Berti_2006,Tattersall:2019nmh,Tattersall:BHspectro,Sirera:2023pbs,Sirera:2024ghv}. Again focusing on the quadrupolar $\ell = 2,n=0$ mode  -- the mode with the largest amplitude for astrophysical mergers \cite{Berti_2006,London:2014cma,Berti:2007fi,Berti:2007zu,Bhagwat:2019bwv,Bhagwat:2019dtm} -- we employ an idealised setup where $\approxcon$ and $M$ are the only free parameters to be determined, while all other waveform parameters $(A,\phi^+,...)$ are assumed to be known. The error on $\approxcon$, $\sigma_\beta$, then satisfies the standard propagation of errors
\begin{align}
    \left(\frac{\sigma_\approxcon}{\approxcon}\right)^2=\left(\frac{\sigma_{M^2/\approxcon}}{M^2/\approxcon}\right)^2+\left(\frac{2\sigma_M}{M}\right)^2,
    \label{eq:error-beta}
\end{align}
where the combination $M^2/\approxcon$ is measured during ringdown, as in \eqref{eq:qnm}, and prior information on $M$ (e.g. from the inspiral or merger) is used to break the degeneracy between $M$ and $\beta$ -- see Appendix~\ref{app:fisher} for details. Defining the real (oscillation frequency) and imaginary (damping time) parts of the QNMs as $\omega=2\pi f-\frac{i}{\tau}$, the combined error is given by
\begin{align}
\sigma_{M^2/\approxcon}\rho = \left|\frac{f}{\sqrt{2}Qf'}\right| \approx 0.67\frac{M^2}{\approxcon},
\end{align}
where the prime denotes a derivative with respect to $(M^2/\approxcon)$, $\rho$ is the ringdown SNR, $Q=\pi f\tau$ is the quality factor, and the numerical result is obtained for the $\ell=2$ mode from the QNM prediction in Eq.~\eqref{eq:qnm}.

\begin{figure}[t!]
    \centering
    \includegraphics[width=\linewidth]{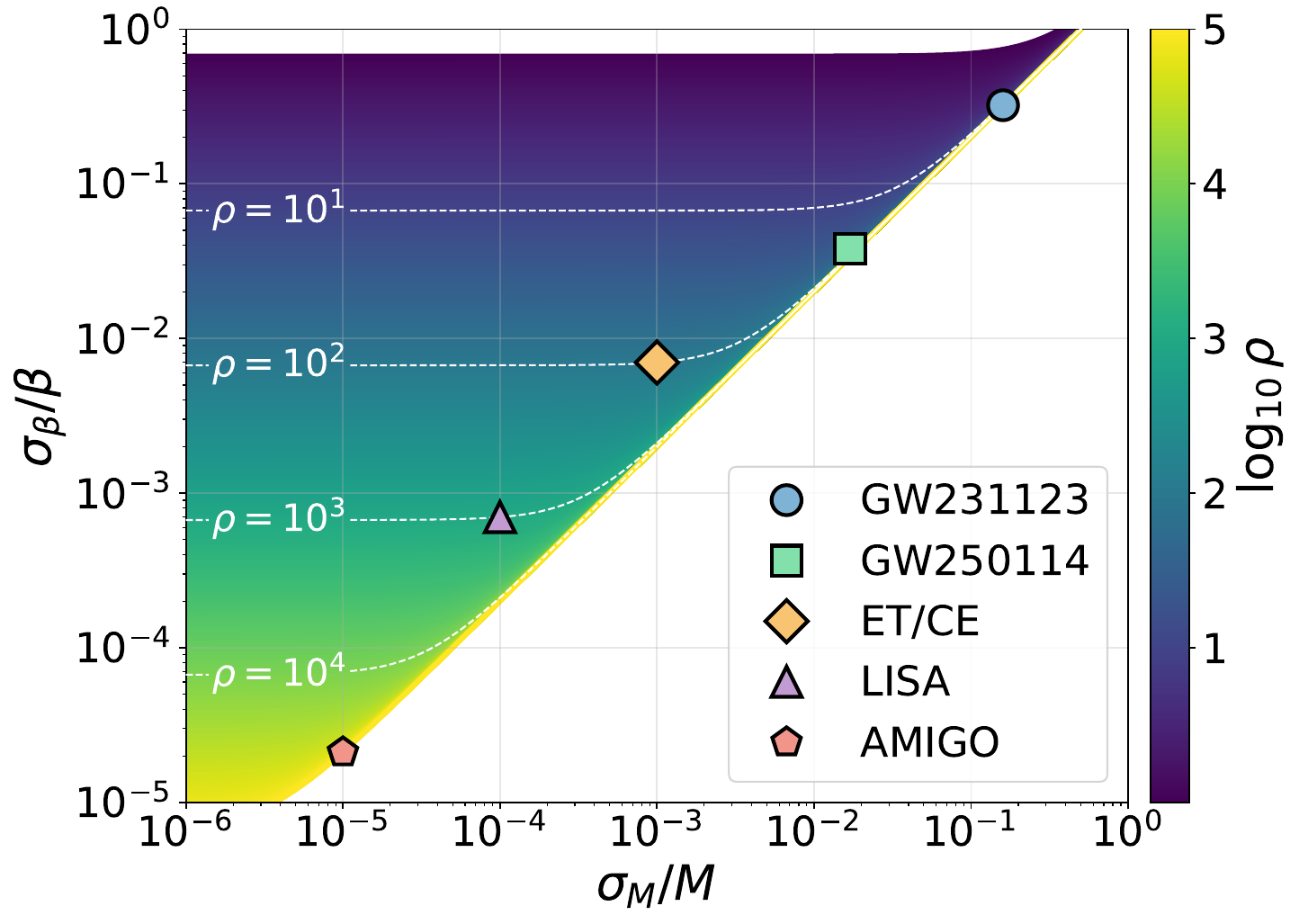}
    \caption{
    Forecasted constraints on the dark energy-induced black hole hair parameter $\approxcon$ as a function of the uncertainty in the inspiral/merger-inferred remnant mass $\sigma_M/M$ and the ringdown SNR $\rho$ -- see Eq.~\eqref{eq:error-beta}. We show the estimated precision of current LVK and forecasted future measurements.
    }
    \label{fig:errors}
\end{figure}

Within the current LVK catalog, GW250114 stands out as the most precisely measured event, with a fractional total mass uncertainty of $(\sigma_M/M)\approx1.7\times10^{-2}$ due to its high SNR $\rho\approx76$ (with $\rho\approx40$ in the ringdown) \cite{LIGOScientific:2025rid,LIGOScientific:2025wao}.
As another example, GW231123 (the most massive event in GWTC-4) exhibits more typical catalog performance with $(\sigma_M/M)\approx1.6\times10^{-1}$ and $\rho\approx20$ (with $\rho\approx18$ in the ringdown) \cite{LIGOScientific:2025slb,Siegel:2025xgb}.
Looking ahead, for next-generation ground based detectors -- Einstein Telescope \cite{ET:2019dnz} (ET) and Cosmic Explorer \cite{Evans:2021gyd} (CE) -- the fractional mass uncertainty can be as good as $10^{-5}$ for `golden events', while the ringdown SNR is expected to reach $\rho \sim {\cal O}(10^2)$ ~\cite{Iacovelli:2022bbs}. For next-generation space-based detectors, specifically LISA \cite{LISA:2022kgy}, ringdown SNR values of $\rho \sim 10^3$ will be reachable \cite{Flanagan:1997sx}, while a fractional mass uncertainty of $(\sigma_M/M) \sim 10^{-4}$ can be obtained for such events, see e.g. the Mock LISA Data Challenge \cite{MockLISADataChallengeTaskForce:2009wir}. Looking further ahead, the proposed AMIGO detector could achieve a ringdown SNR of $\sim 10^5$, together with a fractional mass uncertainty of $(\sigma_M/M) \sim 10^{-5}$  \cite{Baibhav:2019rsa}. The estimates in this paragraph, as well as associated systematics, are detailed in Appendix \ref{app:massprior} and we show the resulting $\sigma_\beta/\beta$ constraints in figure \ref{fig:errors}. Note that, when computing inspiral/merger constraints for the solutions considered here, we (optimistically) assume that mass uncertainties will eventually be obtainable with similar precision as the GR values quoted here. It is worth highlighting, that dark-energy motivated models with an interaction scale $\Lambda_3^3 = \Mpl H_0^2$ (as discussed here) have a naive cutoff at around ${\cal O}(10^2)$ Hz \cite{deRham:2018red}, so firmly within the LVK band. Extracting constraints on e.g. $\beta$ using LVK measurements would therefore implicitly assume that a (low-energy) dark energy theory like \eqref{eq:CubicGalileonAction} is still sufficiently accurate at those scales, i.e. that the UV completion of such theories only softly modifies predictions in this band. This is a highly non-trivial assumption, so predictions for the LISA band are more robust from this perspective (since they are further below the naive cutoff, so less likely to be affected by details of the UV completion).

Constraints on $\approxcon$ can be translated into constraints on $q/q_0$, and hence on the dark energy profile \eqref{eq:Ansatz}, by using Eq.~\eqref{eq:BetaFittingFormula}. Doing so for the $\beta$ bounds highlighted in Figure~\ref{fig:errors}, we find constraints on the linear time dependence of the dark energy field $\sigma_{q/q_0}\sim10^{-2}$ and $\sigma_{q/q_0}\sim10^{-4}$ for LVK and LISA, respectively. Similarly, we can constrain the radial profile $\Psi(t,r)$ of the dark energy field near the black hole using the short-range solution \eqref{eq:ApproximateSolutions}. Note that the precision with which the scalar profile is known will be dominated by the measurement errors in $\approxcon$ and $r_s\propto M$ outlined above, since (for a given $\beta$ and $M$) the solution \eqref{eq:ApproximateSolutions} has a (sub-dominant) error of $\order{Hr_s}$. Finally, to determine the (normalisation-dependent) value of $q$ directly appearing in the ansatz \eqref{eq:Ansatz}, an independent measurement of $H_0$ is required -- see Eq. \eqref{eq:qBound}.\footnote{
Typical errors on $H_0$ from cosmological measurements are currently at the ${\cal O}(1\%)$ level \cite{Planck:2018vyg,Riess:2021jrx,DESI:2024mwx}, and set to improve to the ${\cal O}(0.1\%)$ level with next-generation surveys \cite{Euclid:2021qvm}, so such an uncertainty would in fact dominate the error in $q$ for LISA measurements. Of course we are also working with a simplified long-distance de Sitter limit here, so further work is required to precisely model how this would be modified for cosmological FRW asymptotes.
}

\section{Conclusions} \label{sec:Conclusions}

Testing dark energy with black hole ringdown has long seemed impractical, with a cosmological constant only imprinting minute modifications to the QNM spectra, e.g. ${\cal O}(10^{-34})$ for a black hole of $10^6M_\odot$ \cite{Barausse:2014tra,Sirera:2024ghv}. Meanwhile, known static solutions in dark energy-related theories indicate that no observable QNM shifts can be generated by the interactions driving dark energy dynamics around bald \cite{Tattersall:2019pvx} or hairy \cite{Noller:2019chl} configurations.\footnote{
Note, however, that interactions suppressed on cosmological scales {\it can} lead to observable QNM shifts in such theories \cite{Noller:2019chl}.
}
Known time-dependent solutions, fully capturing this essential feature of dark energy fields, were typically found to be unstable \cite{Babichev:2013cya,Kobayashi:2014eva,Babichev:2016kdt,BenAchour:2018dap,Motohashi:2019sen,Charmousis:2019vnf,deRham:2019gha,Takahashi:2020hso,Khoury:2020aya,Takahashi:2021bml}. All combined, this made it seem questionable whether black hole ringdown would ever provide an insightful laboratory for testing dark energy. However, using the recently identified stable, time-dependent and dynamical dark energy driven black hole solution in \cite{Smulders:2026qwc}, we have here shown how to parametrically connect cosmological and strong field limits of this solution. This implies the dark energy field can imprint observable deviations onto the QNM spectrum, shifting frequencies by up to several tens of percent, and hence opening the door for using ringdown measurements as a novel tool for constraining dark energy. Going forward, more accurate modelling (e.g. better merger modelling to extract more accurate mass estimates, extending the QNM calculation to spinning black hole remnants, performing full parameter estimations) and larger SNR detections with upcoming surveys (such as LISA and ET/CE) are set to make this a precision tool for investigating the underlying nature of dark energy. 

\section*{Statement of Contribution} 

Quasi-normal mode spectrum calculation (LS); Derivation of the link between cosmological and black hole solutions (LS); Forecast and associated error derivation (SS, JN); Project design (JN); drafting of the manuscript (LS, JN, SS); plot creation (SS, LS); revision and proofreading (JN, SS, LS); supervision (JN); interpretation of results (JN, SS, LS). All authors read and approved the final manuscript.

\section*{Acknowledgments}
We thank Pedro Ferreira, Shinji Mukohyama, Kazufumi Takahashi, and Leonardo Trombetta for useful discussions.
LS is supported by the STFC. JN is supported by an STFC Ernest Rutherford Fellowship (ST/S004572/1). SS is supported by ERC Starting Grant SHADE
(grant no. StG 949572). In deriving the results of this paper, we have used \texttt{xAct}~\cite{xAct} and the \texttt{ringdown calculations} repository~\cite{ringdown-calculations}. For the purpose of open access, the authors have applied a Creative Commons Attribution (CC BY) licence to any Author Accepted Manuscript version arising from this work.
\\

\noindent{\bf Data availability} Supporting research data are available on reasonable request from the authors. 

\appendix

\section{\label{sec:AppendixEOMs} Equations of motion}

In order to find numerical solutions, we consider the equations of motion in the quasi-stationary case. A small time dependence is induced by a non-zero radial component of the Noether current due to the shift-symmetry:
\begin{align}
    \label{eq:ConservedCurrent}
    r^2\sqrt{\frac{h}{f}}J^r=\alpha_{BH}\mpl r_s,
\end{align}
where $\alpha_{BH}$ is a constant controlling the time dependence and the full covariant form of the shift-symmetry current is given by
\begin{align}
    \label{eq:ShiftSymmetryCurrent}
    J^\mu
    =2\contwo\partial^\mu\varphi
    +2\frac{\conthree}{\Lambda_3^3}\partial^\mu\varphi\Box\varphi-\frac{\conthree}{\Lambda_3^3}\partial^\mu\left(\partial\varphi\right)^2.
\end{align}
To numerically solve the equations of motion, it is convenient to express them in dimensionless form. To this end we define the following dimensionless parameters:
\begin{align}
    \alpha_1 &\equiv \frac{\conthree}{\contwo}\frac{q}{\Lambda_3^3 r_s},
    &\alpha_2 &\equiv -2\contwo\frac{q^2r_s^2}{\mpl^2},
    \nonumber\\
    \alpha_4 &\equiv \alpha_{BH}\frac{\mpl}{\contwo q r_s}
    & x &\equiv\frac{r}{r_s}.
    \label{eq:DimensionlessParameters}
\end{align}
In terms of these parameters, the quasi-stationary equations of motion are
\begin{align}
    \label{eq:EoMsQuasiStationary}
    \alpha_1(x^4h)'\frac{f}{h}\Xi^2+2x^4h\Xi-\alpha_1x^4h'&=\alpha_4 x^2h\sqrt{\frac{h}{f}},
    \nonumber\\
    \alpha_2x^2\left[1-\frac{f}{h}\Xi^2\right]+2xfh'+2h(f-1)&=0,
    \nonumber\\
    \left[1-\frac{f}{h}\Xi^2\right]\left[\alpha_2x^2\sqrt{\frac{h}{f}}+\alpha_1\alpha_2\left(x^2\sqrt{\frac{f}{h}}\Xi\right)'\right]&
    \nonumber\\
    -2xh^2\left(\sqrt{\frac{f}{h}}\right)'&=0,
\end{align}
where primes indicate derivatives with respect to $x$. Note that the full equations of motion also include time-dependent terms, but this time dependence is suppressed \cite{Smulders:2026qwc}.

Finally, the dimensionless parameters $\alpha_1$ and $\alpha_2$ can be expressed in terms of the physical parameters $q/q_0$, $H$ and $r_s$ used in the main text as
\begin{align}
    \alpha_1  = \frac{1}{3 H r_s}\frac{q}{q_0},&&
    \alpha_2 = 6 \left(\frac{q}{q_0}\right)^2\left( H r_s\right)^2,
\end{align}
while $\alpha_4$ must satisfy
\begin{align}
    \label{eq:StabilityConstraint}
    \frac{\alpha_4\sqrt{\approxcon}}{\alpha_1}=-1+\epsilon, \quad \text{where} \quad |\epsilon|\lsim(Hr_s)^{2/3}
\end{align}
for stability. Note that due to the smallness of $Hr_s$ this effectively fixes $\alpha_4=-\alpha_1/\sqrt{\approxcon}$, which we have used in the main text. For more details on the derivation of these equations of motion and the stability requirement see \cite{Smulders:2026qwc}.

\section{Regge-Wheeler gauge} \label{sec:AppendixRWGauge}

In the Regge-Wheeler gauge, the perturbations have the forms
\begin{widetext}
\begin{align}
    \label{eq:ReggeWheelerGauge}
    r_s^2\delta g_{\mu\nu}dx^\mu dx^\nu&=
    \sum_{\ell=0}^{\infty}\sum_{m=-\ell}^{\ell}\left[
        h(t,r)dt^2H_{\ell m}^{(0)}(t,r)
        +\frac{dr^2}{f(t,r)}H_{\ell m}^{(2)}(t,r)
    \right]Y_{\ell m}(\theta,\phi),
    \nonumber\\
    &+\sum_{\ell=1}^{\infty}\sum_{m=-\ell}^{\ell}\left[
        2dtdrH_{\ell m}^{(1)}(t,r)
        -2r_sdtd\theta\frac{1}{\sin\theta}h_{\ell m}^{(0)}(t,r)\partial_\phi
        +2r_sdtd\phi\sin\theta h_{\ell m}^{(0)}(t,r)\partial_\theta
    \right]Y_{\ell m}(\theta,\phi),
    \nonumber\\
    &+\sum_{\ell=2}^{\infty}\sum_{m=-\ell}^{\ell}\left[
        r^2d\Omega^2K_{\ell m}(t,r)
        -2r_sdrd\theta\frac{1}{\sin\theta}h_{\ell m}^{(1)}(t,r)\partial_\phi
        +2r_sdrd\phi\sin\theta h_{\ell m}^{(1)}(t,r)\partial_\theta
    \right]Y_{\ell m}(\theta,\phi),
    \nonumber\\\nonumber\\
    \delta\varphi&=\sum_{\ell=0}^{\infty}\sum_{m=-\ell}^{\ell}\delta\varphi_{\ell m}(t,r)Y_{\ell m}(\theta,\phi).
\end{align}
\end{widetext}
Here, $h_{\ell m}^{(0)}$ and $h_{\ell m}^{(1)}$ are fields describing metric perturbations of odd parity, while $H_{\ell m}^{(0)}$, $H_{\ell m}^{(1)}$, $H_{\ell m}^{(2)}$ and $K_{\ell m}$ describe metric perturbations of even parity. The (even parity) scalar perturbations are (with a slight abuse of notation) described by the fields $\delta\varphi_{\ell m}$. $Y_{\ell m}$ are the usual spherical harmonic functions.

The two spin-2 degrees of freedom are conveniently expressed in terms of the summary variables
\begin{align}
    w_{\ell m}&=r_s\dot{h}_{\ell m}^{(1)}-r_sh_{\ell m}'^{(0)}+\frac{2r_s}{r}h_{\ell m}^{(0)},
    \nonumber\\
    \psi_{\ell m}&=H_{\ell m}^{(2)}+\frac{(\ell+2)(\ell-1)r+3r_s}{2(r-r_s)}K_{\ell m}-rK'_{\ell m},
\end{align}
where dots and primes denote derivatives with respect to $t$ and $r$, respectively. $w_{\ell m}$ describes odd metric perturbations and $\psi_{\ell m}$ describes even metric perturbations.

\section{Modified Regge-Wheeler and Zerilli equations} \label{app:RWZ}
In this Appendix, we provide some more detail on the derivations of the modified Regge-Wheeler and Zerilli equations \eqref{eq:ModifiedRWZ}. We will use the odd part of the quadratic action as an example, but the derivation for the even metric modes will be analogous, since the quadratic action has the same general form.

The odd part of the quadratic action \eqref{eq:QuadraticActionsFull} has the form
\begin{align}
    \mathcal L_{\text{odd/even}}^{(2)}=\gamma_1(r)\dot{w}^2-\gamma_2(r)w'^2-\gamma_3(r)w^2,
\end{align}
where dots and primes indicate derivatives with respect to the  coordinates $t$ and $r$ respectively. The equation of motion we find is then
\begin{align}
    \ddot{w}-\frac{\gamma_2}{\gamma_1}w''-\frac{\gamma_2'}{\gamma_1}w'+\frac{\gamma_3}{\gamma_1}w=0.
\end{align}
To get the equation in a Schr\"odinger-like form we define a tortoise coordinate $dr_*=dr/A(r)$, with $A(r)\propto\sqrt{\gamma_2/\gamma_1}$. This function is given in section \ref{sec:QNMs}. For both the even and odd modes we are able to choose the proportionality constant such that we obtain exactly the same tortoise function as in GR. We then have an equation of the form
\begin{align}
    \left(\frac{\partial_t^2}{\approxcon}-\partial_*^2\right)w-b_1\partial_*w-b_2w=0,
\end{align}
where $\partial_*=\partial/\partial r_*$. To remove the first order derivative term we define the Regge-Wheeler function $Q(t,r)=B(r)w(t,r)$. We find
\begin{align}
    \left(\frac{\partial_t^2}{\approxcon}-\partial_*^2\right)Q
    &=B\left(\frac{\partial_t^2}{\approxcon}-\partial_*^2\right)w-2\partial_*B\partial_*w-\partial_*^2Bw
    \nonumber\\
    &=\left(Bb_1-2\partial_*B\right)\partial_*w+\left(Bb_2-\partial_*^2B\right)w.
\end{align}
We then require
\begin{align}
    &Bb_1-2\partial_*B=0,
    \nonumber\\
    \Rightarrow &B(r)\propto\exp\int dr_*\frac{b_1(r)}{2}   
    \propto\exp\int dx\frac{b_1(r)}{2A(r)},
\end{align}
which gives $B(r)\propto r$ in the odd case. We then identify
\begin{align}
    B(r)V_{\ell}^{RW}(r)=\partial_*^2B-Bb_2=B\left(\frac{Ab_1'}{2}+\frac{b_1^2}{4}-b_2\right),
\end{align}
which gives
\begin{align}
    V_\ell^{RW}(r)=\left(1-\frac{r_s}{r}\right)\left[\frac{\ell(\ell+1)}{r^2}-\frac{3r_s}{r^3}\right]
\end{align}
for the Regge-Wheeler potential. Similarly, we find for the Zerilli potential
\begin{align}
    V_\ell^Z(r)&=\left(1-\frac{r_s}{r}\right)\frac{8\lambda^2(\lambda+1)r^3+12\lambda^2r_sr^2+18\lambda r_s^2r+9r_s^3}{r^3(2\lambda r+3r_s)^2},
\end{align}
where $\lambda=\left(l+2\right)\left(l-1\right)/2$. Note that these potentials are identical to those in GR.

\section{Fisher information analysis} \label{app:fisher}

In order to derive forecasted constraints on $\approxcon$ we have employed a simplified Fisher information analysis, where only the combination of parameters $M^2/\approxcon$ is to be determined from the ringdown signal, while all other waveform parameters $(A,\phi^+,...)$ are assumed to be known. Details and explicit checks of these assumptions can be found in \cite{Berti_2006}.

Error estimates can be derived from the Fisher information matrix, defined as the noise-weighted inner product
\begin{equation}
    \Gamma_{ab}=\left(\frac{\delta h}{\delta\theta^a}\Big|\frac{\delta h}{\delta\theta^b}\right),
    \label{eq:FM}
\end{equation}
with $h$ being the characteristic ringdown waveform and $\theta^a$ corresponding to the set of parameters to be determined.
Parameter errors can then be obtained by inverting the matrix (giving the covariance matrix $\Sigma$) and taking the square-root of the diagonal elements as
\begin{equation}
    \sigma_a=\sqrt{\Sigma_{aa}}=\sqrt{\Gamma_{aa}^{-1}}.
\end{equation}

\begin{figure}[t!]
    \centering
    \includegraphics[width=\linewidth]{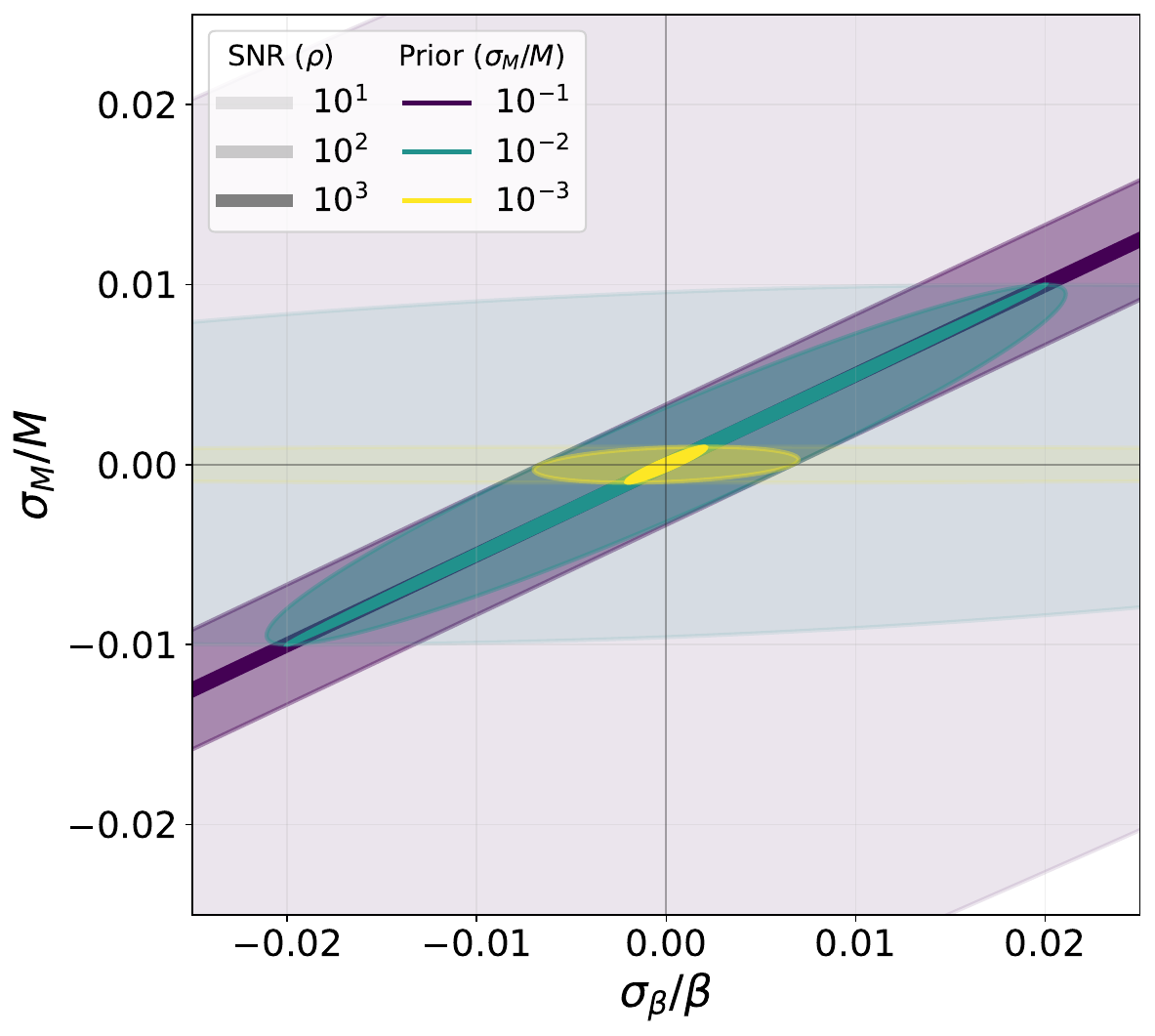}
    \caption{Fisher ellipses in the $(M, \approxcon)$ plane, illustrating the complete degeneracy between the two. Independent prior knowledge on $M$ can effectively break the degeneracy, with the final constraint limited by the SNR of the ringdown signal.}
    \label{fig:fisher-ellipses}
\end{figure}

In the main text, we have shown how $M$ and $\approxcon$ are completely degenerate parameters in the context of ringdown. Hence, with $M^2/\approxcon$ being the only independent parameter combination in the ringdown, its error from the $1$-dimensional Fisher matrix corresponding to the $\ell=2$ mode is given by
\begin{align}
\sigma_{M^2/\approxcon}\rho = \frac{f}{\sqrt{2}Qf'} \approx 0.67\frac{M^2}{\approxcon}.
\label{eq:sigma-M2beta}
\end{align}
Assuming an independent mass measurement, e.g. as determined from the inspiral-merger, we have isolated the error on $\approxcon$ by introducing a mass prior as
\begin{align}
    \left(\frac{\sigma_\approxcon}{\approxcon}\right)^2=\underbrace{\left(\frac{\sigma_{M^2/\approxcon}}{M^2/\approxcon}\right)^2}_{\text{Ringdown}}+\underbrace{\left(\frac{2\sigma_M}{M}\right)^2}_{\text{Prior}}.
    \label{eq:errors}
\end{align}
This expression is illustrated in Figure~\ref{fig:errors}, showing the achievable constraints on $\approxcon$ for some given SNR and mass prior. Additionally, we illustrate in Figure~\ref{fig:fisher-ellipses} the degeneracy between $M$ and $\approxcon$, and how it can be broken by a combination of sufficiently high SNR and precise mass prior, providing tight $\approxcon$ constraints. What this implies for the forecasted precision with which $\approxcon$, and hence the dark energy profile, can be determined for different detectors, is discussed in appendix \ref{app:massprior} and summarised in Table~\ref{tab-SNRringdown}.

Below, we present an alternative derivation of the same expression. Treating $M$ and $\approxcon$ as separate parameters, the total Fisher matrix is constructed as the sum of the ringdown information and the independent mass prior as
\begin{gather}
    \Gamma_T = \Gamma_R + \Gamma_P, \nonumber\\[6pt]
    \Gamma_R = \begin{pmatrix}
    \Gamma_{\approxcon\approxcon} & \Gamma_{\approxcon M} \\
    \Gamma_{\approxcon M} & \Gamma_{MM}
    \end{pmatrix}, \quad 
    \Gamma_P = \begin{pmatrix}
    0 & 0 \\
    0 & 1/\sigma_M^2
    \end{pmatrix},
\end{gather}
Inverting the total matrix $\Gamma_T$ yields the error for $\beta$\footnote{
Note that in order to invert $\Gamma_T$ we have used the fact that $\det(\Gamma_R)=0$, due to the complete degeneracy between $\approxcon$ and $M$.
}
\begin{align}
\sigma_\approxcon^2 = (\Gamma_T^{-1})_{\approxcon\approxcon} = \frac{1}{\Gamma_{\approxcon\approxcon}} + \frac{\Gamma_{MM}\sigma_M^2}{\Gamma_{\approxcon\approxcon}},
\label{eq:sigma-beta-squared}
\end{align}
The first term on the RHS represents the idealised $1$-dimensional error $\sigma_{\approxcon,1D}^2$, in the case where $M$ is known with absolute precision ($\sigma_M=0$). This is given by
\begin{align}
\sigma_{\approxcon,1D}\rho = \left|\frac{f}{\sqrt{2}Qf'}\right| \approx 0.67\approxcon,
\end{align}
where the prime now denotes a derivative with respect to $\approxcon$. Unsurprisingly, this result is identical to Eq.~\eqref{eq:sigma-M2beta}, confirming that the uncertainty in the idealised 1D case is equivalent to the fractional error on the combined ringdown observable, i.e., $\frac{\sigma_{\beta,1\text{D}}}{\beta} = \frac{\sigma_{M^2/\beta}}{M^2/\beta}$.

For the second term on the RHS of Eq.~\eqref{eq:sigma-beta-squared}, given the scaling $\omega \propto \sqrt{\beta}/M$, the Fisher elements can be related via the chain rule, giving $\Gamma_{MM} = (2\beta/M)^2 \Gamma_{\beta\beta}$. Substituting these into Eq.~\eqref{eq:sigma-beta-squared} leads to the same previously derived fractional error
\begin{align}
\left(\frac{\sigma_\beta}{\beta}\right)^2 = \left(\frac{\sigma_{\beta,1D}}{\beta}\right)^2 + \left(\frac{2\sigma_M}{M}\right)^2.
\end{align}

\section{Mass priors and dark energy constraints}\label{app:massprior}

\begin{table}[htp]
\centering
\begin{tblr}{
  colspec = {|c||c|c|c||c|},
  row{1} = {fg=black, font=\bfseries},
  row{2} = {bg=gray9, fg=black},
  row{3} = {bg=gray9, fg=black}
}
\hline[1pt]
   Detector(s) & SNR${}^{(R)}$ ($\rho$) & $\frac{\sigma_{M^2/\beta}^{(R)}}{M^2/\beta}$ & $\frac{\sigma^{(IM)}_{M}}{M}$ & $\frac{\sigma_{\beta}}{\beta}$\\
   \hline[1pt]\hline[1pt]
    LVK & $10$ \cite{TheLIGOScientific:2016src,Flanagan:1997sx,Nakano:2021bbw} & $10^{-1}$ & {$10^{-2}$} & {$10^{-1}$}\\
   \hline[1pt]
       ET / CE & $10^2$ \cite{Maggiore:2019uih,Nakano:2021bbw,Evans:2021gyd,Hall:2022dik}  & $10^{-2}$ & {$10^{-5}$} & {$10^{-2}$}\\
    \hline[1pt] \hline[1pt]
    LISA & $10^3$ \cite{Bhagwat:2021kwv,Yi:2024elj,Shi:2019hqa,Shi:2024ttu} & $10^{-3}$ & {$10^{-4}$} & {$10^{-3}$}\\
   \hline[1pt]
    AMIGO & $10^5$ \cite{Baibhav:2019rsa} & $10^{-5}$ & {$10^{-5}$} & {$10^{-5}$}\\
   \hline[1pt]
\end{tblr}
\caption[ringdownSNR]{
Here we summarise the forecasted precision with which the dark energy parameter $\beta$ -- see the fitting formula \eqref{eq:BetaFittingFormula} -- can be measured for different detectors. We detail (all given as order of magnitude estimates): 1) the actual/predicted SNR achievable for ringdown for these different detectors, 2) the forecasted precision for the ringdown measurement of $M^2/\beta$ -- see Eq.~\eqref{eq:qnm} for why this is the combination measured by ringdown, 3) the forecasted precision of priors on the final ringdown mass from the inspiral and merger phases -- this is used to break the degeneracy between $M$ and $\beta$ in the ringdown measurement, 4) the forecasted overall error on $\beta$, determined using the previous two errors via \eqref{eq:errors}.
Finally, the grey LVK and ET/CE rows are a reminder of the point discussed in Section~\ref{sec:Forecasts}: that the naive cutoff of dark energy models as discussed here lies within the LVK band \cite{deRham:2018red} and therefore extracting dark energy parameter constraints using LVK-band measurements requires additional assumptions on the UV completion of such theories (while LISA band measurements are more robust from this perspective).}
    \label{tab-SNRringdown}
\end{table}

In Section~\ref{sec:Forecasts}, specifically around Eq.~\eqref{eq:error-beta}, we discussed how an additional, non-ringdown measurement of the mass of the remnant black hole $M$ is required to break the degeneracy between $M$ and $\beta$, i.e. to obtain a measurement of $\beta$ itself. Inferring the remnant black hole mass $M$ from the inspiral/merger part of the signal is a key part of so-called inspiral--merger--ringdown consistency tests, that compare this prediction with the ringdown-only measurement of the same mass -- see~\cite{Hughes:2004vw,Ghosh:2016qgn,Ghosh:2017gfp} as well as~\cite{LIGOScientific:2026qni} for a recent application. Note that, as in the main text, we are labelling the remnant black hole mass as $M$ and will refer to the total mass of the system as $M_t$. 
The chirp mass $\mathcal{M}_c$ and symmetric mass ratio $\eta$ are determined from the inspiral/merger part of the signal, from which the total mass $M_t = \mathcal{M}_c\,\eta^{-3/5}$ follows. The remnant mass is then predicted as
\begin{equation}
  M = f_{\rm NR} \cdot M_t,
  \qquad
  f_{\rm NR} \equiv 1 - \frac{E_{\rm rad}}{M_t},
  \label{eq:mf_prediction}
\end{equation}
where $f_{\rm NR}$ is a dimensionless function inferred from numerical-relativity (NR)
simulations (that depends on the mass ratio $q$ of the binary system as well as component spin vectors).
$E_{\rm rad}$ is the total mass-energy radiated in gravitational waves, $E_{\rm rad} = M_t - M$. Key to the inspiral--ringdown mapping is an accurate modelling of the merger, encoded in $f_{\rm NR}$ (or, equivalently, in $E_{\rm rad}$). 

The key factors setting the accuracy with which the remnant mass $M$ can be determined in this way are therefore 1) the measurement precision for ${\cal M}_c$ and $\eta$, and 2) the accuracy with which $f_{\rm NR}$ can be computed. 
The precision of ${\cal M}_c$ and $\eta$ measurements depends on survey characteristics, as discussed in detail below.
The $f_{\rm NR}$ mapping, on the other hand, can e.g. be obtained from the \textsc{surfinBH}/\textsc{NRSur} family of surrogate remnant models~\cite{Varma:2018aht,vijay_varma_2018_1435832,Varma:2019csw,Boschini:2023ryi,MaganaZertuche:2024ajz} 
resulting in an estimate of $M$ (for a given ${\cal M}_c$ and $\eta$) accurate at the $\sim 10^{-4}$ level \cite{MaganaZertuche:2024ajz}. 
This current systematic modelling floor is already below the benchmark (SNR-induced) ringdown uncertainty $\sigma_{M^2/\beta}/(M^2/\beta) \approx 6.7 \times 10^{-4}$ at $\rho \sim 10^3$ relevant for LISA, so the expected accuracies for next-generation surveys quoted in  Table~\ref{tab-SNRringdown} would already be realisable with such modelling accuracies. Note, however, that these surrogate models are built in the context of standard GR waveforms and adapting this to beyond-GR solutions as considered here (e.g. building on the work of  \cite{deRham:2012fw,Dar:2018dra,Brax:2020ujo,Figueras:2021abd}) would of course be a very significant task -- in the same vein, we are here (optimistically) assuming that mass uncertainties will eventually be obtainable with similar precision as the GR values quoted here.

For current LVK detectors, GW250114~\cite{LIGOScientific:2025rid,LIGOScientific:2025wao} serves as a good example of the associated constraining power, with a large SNR $\rho\approx76$ (with $\rho\approx40$ in the ringdown). This high mass ($M_t\approx66\,M_\odot$) system provides exquisite constraints of the ringdown phase, while the inspiral/merger-inferred $M$ is accurate at the percent level (comparable to its ringdown determination). More specifically, ${\cal M}_c$ is constrained to a fractional precision of $\sim1.7\%$ and the system is bounded to be nearly equal-mass \cite{LIGOScientific:2025rid}, which tightly restricts the symmetric mass ratio and translates to an inspiral/merger-inferred remnant with $\sim 1.7\%$ accuracy.
For next-generation ground based detectors -- Einstein Telescope \cite{ET:2019dnz} (ET) and Cosmic Explorer \cite{Evans:2021gyd} (CE) -- `golden events' can reach a ringdown SNR of $\rho \sim {\cal O}(10^2)$, while the measurement precision for ${\cal M}_c$ and $\eta$ can reach levels of ${\cal O}(10^{-5})$ -- see~\cite{Iacovelli:2022bbs}.
Note that reaching both accuracies with a single event can be challenging (and would require e.g. a golden GW250114-like observation with ET/CE), since events that are optimal in terms of constraints on ${\cal M}_c$ and $\eta$ are lighter mergers (where a large number of inspiral cycles and the merger are captured at frequencies with optimal sensitivity), whereas optimising the ringdown SNR requires higher masses (to move the ringdown into the optimal frequency sensitivity band).  
However, as can be seen in Figure~\ref{fig:errors}, the limiting factor for constraints on $\beta$ will be the (ringdown) SNR for ET/CE -- even where the measurement accuracy ${\cal M}_c$ and $\eta$ degrades by up to three orders of magnitude from their optimal values -- and we therefore expect $\sigma_\beta/\beta \sim {\cal O}(10^{-2})$.\footnote{Note that we loosely list the best-case ${\cal O}(10^{-5})$ error on the remnant mass prior from the inspiral and merger parts of the signal in Table~\ref{tab-SNRringdown}, but the subtleties discussed here should be understood when interpreting this value. If an inspiral/merger estimate of $M$ accurate at the $10^{-5}$ level can indeed be achieved in principle, this would also require improvements in numerical modelling or, alternatively, this accuracy would be reduced e.g. to the $10^{-4}$ level with numerical modelling uncertainties as in \cite{MaganaZertuche:2024ajz}.
}
In the case of LISA, roughly $\mathcal{O}(10^2)$ events will be detected with errors better than $1\%$ for both individual masses at a threshold of $\rho=8$ \cite{Klein:2015hvg}. Out of those, given the standard scaling $\sigma_M/M \propto \rho^{-1}$ (where here $\rho$ refers to the inspiral/merger SNR), we can expect `best-case' events at high SNR ($\rho \sim 10^3$) to reach accuracies of $(\sigma_M/M) \sim 10^{-4}$.\footnote{Note that for SMBHB both the inspiral and ringdown phases can simultaneously reach $10^3$ SNRs in the LISA band \cite{Flanagan:1997sx}.} Indeed, the Mock LISA Data Challenge \cite{MockLISADataChallengeTaskForce:2009wir} predicts that for such high-SNR sources, the (redshifted) chirp mass $\mathcal{M}_c$ can be recovered with a fractional precision of $10^{-5}$ and the symmetric mass ratio $\eta$ to $10^{-4}$. Recalling that the total mass is given by $M=\mathcal{M}_c\eta^{-3/5}$, its uncertainty is dominated by $\eta$, leading to a comparable total and remnant mass precision of $(\sigma_M/M) \sim 10^{-4}$.
The resulting forecasted constraints on $\approxcon$ are presented in Figure~\ref{fig:errors}.
Finally, looking even further into the future, the proposed AMIGO detector \cite{Baibhav:2019rsa} would provide observations in $\sim$ the same frequency bands as LISA, but with an order of magnitude improvement in sensitivity throughout. 
Note that even the $(220)$ ringdown mode alone could achieve an SNR $\sim 10^5$ for AMIGO for `golden' events with remnant masses of few $\times 10^6 \; M_\odot$ \cite{Baibhav:2019rsa}. For this mass range, AMIGO could also measure ${\cal M}_c$ and $\eta$ with accuracies of orders $10^{-7}$ and $10^{-5}$, respectively \cite{Berti:2004bd}, so a remnant mass prior of order $(\sigma_M/M) \sim 10^{-5}$ should be achievable (with the same numerical modelling caveat discussed above).
Altogether, this implies a highly accurate $\sigma_\beta/\beta \sim {\cal O}(10^{-5})$ would be obtainable with AMIGO. Interestingly, for all other detectors the measurement accuracy for $\beta$ was ultimately limited by ringdown SNR and the associated measurement of $M^2/\beta$, but for AMIGO the measurement precision for this parameter combination and for $\eta$ become comparable. The overall constraint levels for all detectors discussed are summarised in Table~\ref{tab-SNRringdown}.

\bibliographystyle{utphys}
\bibliography{TestDE_QNM}
\end{document}